\documentclass[%
 reprint,
 amsmath,amssymb,
 aps,
]{revtex4-2}

\usepackage{pkgfile}
\usepackage{xifthen}
\usepackage{tabularx}
\usepackage{adjustbox}
\usepackage{pgfplots}
\usepackage{bbm}
\usepackage{makecell}

\usepackage{tikz}
\usetikzlibrary{matrix,arrows}

\usepackage{graphicx}
\usepackage{dcolumn}
\usepackage{bm}

\begin{document}


	\title[Statistical Cooper Pair] {The Universal Gap-to-Critical Temperature Ratio in Superconductors: a Statistical Mechanical Perspective}

\author{Chung-Ru Lee}
\email{math.crlee@gmail.com}

\date{\today}
\counterwithout{equation}{section}

\begin{abstract}
We propose a statistical mechanical framework to unify the observed relationship between the superconducting energy gap $\Delta$, the pseudogap $\Delta^\ast$, and the critical temperature $T_\mathrm{c}$. In this model, fermions couple as a composite boson and condense to occupy a single bound state as the temperature drops. We derive a concise formula for $T_\mathrm{c}$ in terms of $\Delta$ and $\Delta^\ast$, namely:
$$\frac{\Delta}{k_\mathrm{B}T_\mathrm{c}}=1.4+4\log(\Delta^\ast/\Delta).$$
This expression reproduces the standard BCS gap-to-\( T_\mathrm{c} \) ratio in the absence of a pseudogap, while naturally explaining its enhancement in unconventional superconductors. The model is supported by comparisons with experimental data from several cuprates and iron-based superconductors, which highlight its generality. This formulation also offers a theoretical explanation for the observed persistence of the pseudogap phase into the overdoped regime.
\end{abstract}

\date{\today}
\maketitle


\section{Introduction}
One of the renowned achievements of the Bardeen–Cooper–Schrieffer (BCS) theory is that it provides a formulation that explains the relation between the gap $\Delta$ from the energy spectrum and the critical temperature $T_\mathrm{c}$. The theory predicts the relation to be a linear one, and found the theoretical gap-to-$T_\mathrm{c}$ ratio for superconductors:
$$\frac{\Delta}{k_\mathrm{B}T_\mathrm{c}}\approx1.76.$$
This result is material-independent. Measurements in many weak-coupling $s$-wave superconductors are mostly consistent with this prediction. Note that $\Delta=\Delta(0)$ is the superconducting gap at zero temperature.


It may come out as a surprise, however, that this ratio (and its implied proportional relation) seems to also modestly apply to several unconventional superconductors like high-temperature superconductors, and even to some topological superconductors.

Those superconducting phenomena are believed to arise from quite different microscopic mechanisms, such as weak coupling, strongly correlated interactions, or topological interactions. It led us to wonder if there exists an explanation behind this universality.

The objective of this article is threefold:
\begin{enumerate}
\item We want to provide a statistical mechanical point of view on why there seems to exist a universal gap-to-$T_\mathrm{c}$ ratio that serves as a baseline among the various superconductors.
\item Experimental data for high-temperature superconductors, especially the cuprate superconductors indicates that the ratio $\Delta/k_\mathrm{B}T_\mathrm{c}$ is often significantly higher than the prediction of BCS theory. We will derive a formula (\ref{eq:main}) that explains this phenomenon, and predicts the higher ratios to a great precision. The derivation of the formula incorporates the pseudogap phase that has been widely observed in high-temperature superconductors. In particular, this formula involves the ratio $\Delta^\ast/\Delta$, where $\Delta^\ast$ is the pseudogap. In particular, the introduction of the pseudogap $\Delta^\ast$ explains why in many underdoped cuprate superconductors, as the doping level raises, $T_\mathrm{c}$ might increase even though $\Delta$ is decreasing.
\item An observation from experiments is that superconductivity turns out to be difficult to create in materials with a high conductivity at room temperature, such as metals. This article will also try to give a theoretical answer to it.
\end{enumerate}

In some high-temperature superconductors, a longstanding puzzle is the apparent persistence of the pseudogap phase well into the overdoped regime, beyond the point where it is customarily expected to vanish. As an application of this study, we offer a straightforward explanation for this phenomenon.

\section{Formulation}\label{sec:intro}
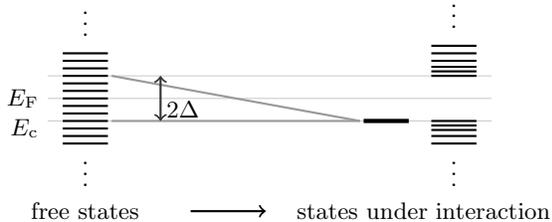
\begin{figure}
    \centering
\begin{tikzpicture}[thick]
    \draw[thin,gray!40] (-2.5,0.8) -- (3.4,0.8);
    \draw[thin,gray!40] (-2.5,0.2) -- (3.4,0.2);
    \draw[thin,gray!40] (-2.5,0.5) -- (3.4,0.5);

    \foreach \y in {-0.1,0,0.1,0.2,0.3,0.4,0.5,0.6,0.7,0.8,0.9,1,1.1} {
        \draw (-2.3,\y) -- (-1.7,\y);
        }

    \foreach \y in {0.8,0.2} {
        \draw[black!40] (-1.65,\y) -- (1.65,0.2);
        }
    \node at (-2,1.6) {$\vdots$};
    \node at (-2,-0.4) {$\vdots$};
    \draw (-2.5,0.5) node[left] {$E_\text{F}$};
    
    \draw[<->,black!80] (-1,0.2) -- (-1,0.8);
    \node at (-0.7,0.35) {$2\Delta$};
    \draw[ultra thick] (1.7,0.2) -- (2.3,0.2);
    \draw (-2.5,0.1) node[left] {$E_{\text{c}}$};
    \foreach \y in {1.2,1.1,1.0,0.92,0.86,0.8,0.2,0.14,0.08,0,-0.1} {
        \draw (2.6,\y) -- (3.2,\y);
        }
    \node at (2.9,1.7) {$\vdots$};
    \node at (2.9,-0.4) {$\vdots$};

    \node at (-2,-1) {free states};
    \node at (2.5,-1) {states under interaction};
    
    \draw[->,thick] (-0.6,-1) -- (0.4,-1);
    \end{tikzpicture}
    \caption{A heuristic for the spectrum of Cooper pairs.}
    \label{fig:feynman}
\end{figure}

We study the statistical properties of a system with the following two characterizing assumptions:
\begin{enumerate}
\item The electrons in states near the Fermi level $E_\mathrm{F}$ degenerate into a single bound state of energy $E_\text{c}$. The electrons in the bound state will be treated pairwise. In regard of the statistics, each pair of coupled electrons behaves as a composite boson \cite{feynman} (see Figure \ref{fig:feynman}).
\item The statistical distribution of electrons (even those slightly) deviate from the Fermi level remains vastly unaffected by the interaction, and the variation in the distribution is negligible.
\end{enumerate}

In addition to the picture in Figure \ref{fig:feynman}, we propose to employ the \textit{distribution for exclusive fermions} \cite{lee,4,5} here, instead of the Fermi-Dirac distribution for free electron gases. In the distribution for exclusive fermions, electrons are prohibited from having double occupancy of a single state, regardless of their spin.

A justification for the choice of such a distribution is that (effective) interaction between electrons is necessary for the formulation of Cooper pairs; therefore, it would be bizarre to apply the Fermi-Dirac distribution here, which describes free electron gases. Furthermore, the impurity of the tested materials seems to manifest itself to be crucial in raising the critical temperature $T_\mathrm{c}$ for superconductors. The distribution of exclusive fermions often becomes the suitable candidate for describing the statistical behavior of the electrons when impurities are involved \cite{4,5}.

For further discussion on the results when using the Fermi-Dirac distribution, see \S\ref{sec:FD}.

Note that at this stage, unlike in many theories for high-temperature superconductors, we do not treat the excited states of the coupled electrons (of energy greater than $E_\mathrm{F}+\Delta$) just as in BCS theory. In such a scenario, the excited electron pairs decouple and return to the fermion spectrum in our statistics, and the pseudogap is absent. We will take in the pseudogap (and therefore the excited states for coupled electrons) later in \S\ref{sec:high}.

The upshot of this article is that from these assumptions and treatments, we derived a statistics-theoretical formula
\begin{align}
\frac{\Delta}{k_\mathrm{B}T_\mathrm{c}}\approx1.4+4\log(\Delta^\ast/\Delta)\label{eq:main}
\end{align}
that relates $\Delta$, $\Delta^\ast$ and $T_\mathrm{c}$. It is worth mentioning that this formula is valid in either the 2-dimensional or 3-dimensional models.

\subsection{The Distribution of Exclusive Fermions}
The statistical distribution of exclusive fermions can be derived by employing the grand canonical ensemble \cite{1}. The $N$-particle canonical partition function $Z_N^F$ can be written as the sum
\begin{align}
\begin{split}
Z_N^F=&Z_N^{\hat{1}}+Z_{N-1}^{\hat{1}}e^{-\beta\epsilon_1^\uparrow}+Z_{N-1}^{\hat{1}}e^{-\beta\epsilon_1^\downarrow}\\
=&Z_N^{\hat{1}}+2Z_{N-1}^{\hat{1}}e^{-\beta\epsilon_1}.
\end{split}
\end{align}
The superscript $\hat{1}$ indicates that $Z_N^{\hat{1}}$ is the $N$-particle canonical partition function for which the state $|1\rangle$ (with energy $\epsilon_1=\epsilon_1^\uparrow=\epsilon_1^\downarrow$) is removed from the spectrum.

The grand canonical partition function $\mathcal{Z}_G^\mathrm{F}$ is then
$$\mathcal{Z}_G^\mathrm{F}=\sum_{N=0}^\infty Z_Nz^N=\prod_i(1+2e^{-\beta\epsilon_i}z).$$
Here, $z=e^{\beta\mu}$ is the fugacity and $\beta=k_\mathrm{B}T$. The final equality is achieved by specifying all the single-particle states $|i\rangle$ inductively.

From this, we yield the expectation for the number of particles
$$\langle N\rangle=z\frac{\partial\log \mathcal{Z}_G^\mathrm{F}}{\partial z}=\sum_i\frac{2}{e^{\beta(\epsilon_i-\mu)}+2}=\sum_i\langle n_i\rangle.$$
Taking the continuum, we have the distribution function of exclusive fermions
\begin{align}
f(\epsilon)=\frac{2}{e^{\beta(\epsilon-\mu)}+2}.\label{eq:ex_dist}
\end{align}

\subsection{The Fermion-Boson Interchanging States}
We proceed with the previous derivation, but with an interaction included.

As suggested in Figure \ref{fig:feynman}, we will allow electrons to pair up with a binding energy $\frac{1}{2}\Delta$. Bosons, and therefore the paired electrons, are unrestricted from multiple occupancy of the single bound state (with condensate energy $E_\mathrm{c}$).


Note that since the distribution function must be non-negative, it is required that the chemical potential $\mu<E_\mathrm{c}$, just as in the derivation of the Bose-Einstein distribution.

In fact, the single bound state remains vacant as $T$ lowers. Electron pairs do not form until $T<T_\mathrm{c}$, a certain critical temperature.

\section{Condensation for Exclusive Fermions}
When the temperature drops to a critical temperature $T_\mathrm{c}$, fermions (on the Fermi surface, gifted energy $E_\mathrm{F}$) may pair up to form bound pairs, with a binding energy $\frac{1}{2}\Delta=E_\mathrm{F}-E_\mathrm{c}$. Such pairs of fermions would then demonstrate bosonic behaviors, among them most importantly the condensation.

\subsection{In Two Dimensions}
We calculate the total number of particles (in this case, electrons as exclusive fermions)
$$N=\int_0^\infty\mathcal{D}(\epsilon)f(\epsilon)d\epsilon.$$
Now the density of states $\mathcal{D}(\epsilon)=\frac{mV}{2\pi\hbar^2}=:A$ in two dimensions is independent of the energy $\epsilon$. When the temperature drops to $T_\mathrm{c}$, the chemical potential $\mu$ would also approach its critical value $E_\mathrm{c}$. Therefore,
\begin{align}
N & =A\int_0^\infty\frac{2}{e^{\beta_\mathrm{c}(\epsilon-E_\mathrm{c})}+2}d\epsilon\notag\\
& =A\left(E_\mathrm{c}+\frac{1}{\beta_\mathrm{c}}\log(e^{-\beta_\mathrm{c}E_\mathrm{c}}+2)\right).\label{eq:2d_critical}
\end{align}

On the other hand, when $T\rightarrow0$, we have
$$N=AE_\mathrm{F}.$$
Combine this and Equation (\ref{eq:2d_critical}) to yield
$$\frac{1}{2}\Delta=E_\mathrm{F}-E_\mathrm{c}=\frac{1}{\beta_\mathrm{c}}\log(e^{-\beta_\mathrm{c}E_\mathrm{c}}+2).$$
Since $e^{-\beta_\mathrm{c}E_\mathrm{c}}\approx0$ at low temperature, we obtain the gap-to-$T_\mathrm{c}$ ratio:
\begin{align}
\Delta=2\log2k_\mathrm{B}T_\mathrm{c}\approx1.4k_\mathrm{B}T_\mathrm{c}.\label{eq:log2}
\end{align}

\subsection{In Three Dimensions}
We plug in the density of states for the $3$-dimensional case
$$\mathcal{D}(\epsilon)=\frac{V}{4\pi^2\hbar^3}(2m)^{3/2}\epsilon^{1/2}=:B\epsilon^{1/2}.$$
Following the method of Sommerfeld expansion at low temperature (that is, when $k_\mathrm{B}T_\mathrm{c}\ll E_\mathrm{c}$), we obtain
\begin{align}
N & =B\int_0^\infty\frac{2\epsilon^{1/2}}{e^{\beta_\mathrm{c}(\epsilon-E_\mathrm{c})}+2}d\epsilon\notag\\
& =\frac{2}{3}B\left(E_\mathrm{c}^{3/2}+3CE_\mathrm{c}^{1/2}(k_\mathrm{B}T_\mathrm{c})\right),\label{eq:old1}
\end{align}
where
$$C=\int_{-\infty}^\infty\frac{xe^x}{(e^x+2)^2}dx\approx0.35.$$

From the relation between particle number and Fermi energy, $N=\frac{2}{3}BE_\mathrm{F}^{3/2}$, we have
$$E_\mathrm{F}\approx E_\mathrm{c}(1+2C\left(\frac{k_\mathrm{B}T_\mathrm{c}}{E_\mathrm{c}}\right)).$$
Thus, the gap-to-$T_\mathrm{c}$ ratio is
\begin{align}
\Delta\approx1.4k_\mathrm{B}T_\mathrm{c}.\label{eq:bcs}
\end{align}

It is a surprise to observe that the universal $\Delta$-$T_\mathrm{c}$ relation remains somewhat consistent across $2$-dimensional and $3$-dimensional systems, and seems insensitive to the dimensionality.

\section{The Pseudogap Phase Adjustment}\label{sec:high}

In many novel high-temperature superconductors, particularly the cuprate superconductors, the experimental data for $\Delta/k_\mathrm{B}T_\mathrm{c}$ are higher than what has been predicted by BCS theory. In this section, we will make an attempt to explain the statistical formulation behind such an elevation in the gap-to-$T_\mathrm{c}$ ratio.

A key feature of this sort of superconductors lies in the presence of a pseudogap. The pseudogap phase is characterized by a partial suppression of the electron density of states near the Fermi level under a certain temperature $T^\ast>T_\mathrm{c}$. It suggests the existence of preformed Cooper pairs that lack long-range phase coherence in such (for example, in underdoped cuprates) cases.

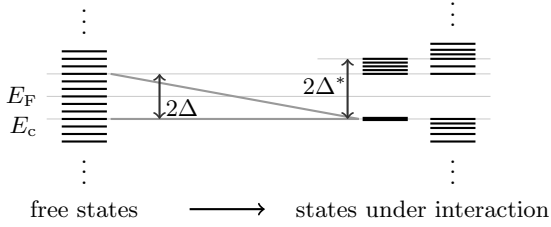
\begin{figure}
    \centering
\begin{tikzpicture}[thick]
    \draw[thin,gray!40] (-2.5,0.8) -- (3.4,0.8);
    \draw[thin,gray!40] (-2.5,0.2) -- (3.4,0.2);
    \draw[thin,gray!40] (-2.5,0.5) -- (3.4,0.5);
    \draw[thin,gray!40] (1.1,1) -- (3.4,1);

    \foreach \y in {-0.1,0,0.1,0.2,0.3,0.4,0.5,0.6,0.7,0.8,0.9,1,1.1} {
        \draw (-2.3,\y) -- (-1.7,\y);
        }

    \foreach \y in {0.8,0.2} {
        \draw[black!40] (-1.65,\y) -- (1.65,0.2);
        }
    \node at (-2,1.6) {$\vdots$};
    \node at (-2,-0.4) {$\vdots$};
    \draw (-2.5,0.5) node[left] {$E_\text{F}$};
    
    \draw[<->,black!80] (-1,0.2) -- (-1,0.8);
    \draw[<->,black!80] (1.5,0.2) -- (1.5,1);    
    \node at (-0.7,0.35) {$2\Delta$};
    \node at (1.2,0.65) {$2\Delta^\ast$};    
    \draw[ultra thick] (1.7,0.2) -- (2.3,0.2);
    \foreach \y in {0.8,0.85,0.9,0.95,1} {
        \draw (1.7,\y) -- (2.3,\y);
        }
    \draw (-2.5,0.1) node[left] {$E_{\text{c}}$};
    \foreach \y in {1.2,1.12,1.06,1.0,0.9,0.8,0.2,0.14,0.08,0,-0.1} {
        \draw (2.6,\y) -- (3.2,\y);
        }
    \node at (2.9,1.7) {$\vdots$};
    \node at (2.9,-0.4) {$\vdots$};

    \node at (-2,-1) {free states};
    \node at (2.5,-1) {states under interaction};
    
    \draw[->,thick] (-0.6,-1) -- (0.4,-1);
    \end{tikzpicture}
    \caption{A heuristic for the spectrum of Cooper pairs with the presence of a pseudogap $\Delta^\ast$.}
    \label{fig:pseudogap}
\end{figure}

To account for this phenomenon, therewithal the previous picture (Figure \ref{fig:feynman}), we incorporate excited states for the paired electrons with energy ranging from $E_\mathrm{c}+2\Delta$ to $E_\mathrm{c}+2\Delta^\ast$ (see Figure \ref{fig:pseudogap}). This extension offers an interpretation of the suppressed electronic density of states as arising from preformed electron pairs with energies up to $2\Delta^\ast$ above the condensate energy $E_\mathrm{c}$. We will focus particularly on the regime when the temperature $T$ is just slightly above $T_\mathrm{c}$, where the superconducting gap is just beginning to open. In this setting, paired electrons are expected to only occupy states outside the fully gapped region $E_\mathrm{c}$ and $E_\mathrm{c}+2\Delta$. As might be expected, we will not impose restrictions on the number of paired electrons occupying such states.

\subsection{The Adjusted Distribution}

It is reasonable to assume that the paired electrons (in their excited states) follow the Bose-Einstein distribution. We then obtain the adjusted distribution function $f^\ast=f^\ast(\epsilon,\mu,T)$, defined by
$$f^\ast=\begin{cases}
\frac{2}{e^{\beta(\epsilon-\mu)}+2}+\frac{2}{e^{\beta(\epsilon-\mu)}-1}&2\Delta\leq\epsilon-E_\mathrm{c}\leq 2\Delta^\ast\\
\frac{2}{e^{\beta(\epsilon-\mu)}+2}&\text{otherwise}
\end{cases}.$$
In brief, we still follow the assumption in Section \ref{sec:intro} that the change in the statistical distribution of fermions remains negligible after the interaction. Thus, $f^\ast(\epsilon)$ is the distribution of $f(\epsilon)$ displayed in Equation (\ref{eq:ex_dist}) with an overlay of the Bose-Einstein distribution over a narrow energy interval. The factor $2$ here accounts for the fact that those are \textit{coupled} electrons in the distribution. Note that $\Delta^\ast\geq\Delta$ by construction.

With this adjusted distribution function, we can study the association between $\Delta$, $T_\mathrm{c}$, and the newly introduced $\Delta^\ast$.

\subsection{In Two Dimensions}
At $T=T_\mathrm{c}$, the total number of particles $N$ is given by (let $x=\epsilon-E_\mathrm{c}$)
\begin{widetext}
\begin{align}
N & =A\left(\int_0^\infty\frac{2}{e^{\beta_\mathrm{c}(\epsilon-E_\mathrm{c})}+2}d\epsilon+\int_{E_\mathrm{c}+2\Delta}^{E_\mathrm{c}+2\Delta^\ast}\frac{2}{e^{\beta_\mathrm{c}(\epsilon-E_\mathrm{c})}-1}d\epsilon\right)\notag\\
& =A\left(E_\mathrm{c}+\frac{1}{\beta_\mathrm{c}}\log(e^{-\beta_\mathrm{c}E_\mathrm{c}}+2)+2\int_{2\Delta}^{2\Delta^\ast}\frac{1}{e^{\beta_cx}-1}dx\right)\notag\\
& =A\left(E_\mathrm{c}+\frac{1}{\beta_\mathrm{c}}\log(e^{-\beta_\mathrm{c}E_\mathrm{c}}+2)+\frac{2}{\beta_c}\log(1-e^{-\beta_cx})|_{x=2\Delta}^{2\Delta^\ast}\right)\label{eq:new1}\\
& \approx A\left(E_\mathrm{c}+\frac{1}{\beta_\mathrm{c}}\log2+\frac{2}{\beta_c}\log({\Delta^\ast}/{\Delta})\right).\label{eq:new2}
\end{align}
\end{widetext}
From (\ref{eq:new1}) to (\ref{eq:new2}) we used the approximation $e^{-\beta_\mathrm{c}E_\mathrm{c}}\approx0$ at low temperature, and
$$\log(1-e^{-\beta_cx})\approx\log(\beta_cx)$$
when $\beta_cx$ is tiny.

As before, when $T\rightarrow0$, we have
$$N=AE_\mathrm{F}.$$
Combine this and Equation (\ref{eq:new2}) to yield
$$\frac{1}{2}\Delta=E_\mathrm{F}-E_\mathrm{c}\approx\frac{1}{\beta_\mathrm{c}}\log2+\frac{2}{\beta_c}\log({\Delta^\ast}/{\Delta})$$
or equivalently,
$$\Delta\approx\left(1.4+4\log({\Delta^\ast}/{\Delta})\right)k_\mathrm{B}T_\mathrm{c}.$$

\subsection{In Three Dimensions}
Notice that  once we plug in the density of states, there is only one term that awaits evaluation in the calculation for $N$ at $T=T_\mathrm{c}$, namely
\begin{align}
\int_{E_\mathrm{c}+2\Delta}^{E_\mathrm{c}+2\Delta^\ast}\frac{2B\epsilon^{1/2}}{e^{\beta_\mathrm{c}(\epsilon-E_\mathrm{c})}-1}d\epsilon\approx
\int_{2\Delta}^{2\Delta^\ast}\frac{2BE_\mathrm{c}^{1/2}}{e^{\beta_\mathrm{c}x}-1}dx.\label{eq:new3}
\end{align}
Since $\beta_cx$ is small, we can approximate (\ref{eq:new3}) by
\begin{align}
\int_{2\Delta}^{2\Delta^\ast}\frac{2BE_\mathrm{c}^{1/2}}{\beta_cx}dx=\frac{2BE_\mathrm{c}^{1/2}}{\beta_c}\log(\Delta^\ast/\Delta).\label{eq:new4}
\end{align}

Compare the sum of Equations (\ref{eq:old1}) and (\ref{eq:new4}) to
$$N=\frac{2}{3}BE_\mathrm{F}^{3/2},$$
we have
$$E_\mathrm{F}^{3/2}=E_\mathrm{c}^{3/2}\left(1+(3C+3\log(\Delta^\ast/\Delta))\frac{k_\mathrm{B}T_\mathrm{c}}{E_\mathrm{c}}\right).$$
Thus
$$E_\mathrm{F}\approx E_\mathrm{c}\left(1+2(C+\log(\Delta^\ast/\Delta))\frac{k_\mathrm{B}T_\mathrm{c}}{E_\mathrm{c}}\right)$$
and so
$$\Delta\approx\left(1.4+4\log({\Delta^\ast}/{\Delta})\right)k_\mathrm{B}T_\mathrm{c}.$$

It is again quite consistent across dimensionalities of $2$ and $3$. We therefore obtain the universal $\Delta$-$T_\mathrm{c}$ formula in Equation (\ref{eq:main}).

We now examine Equation (\ref{eq:main}) against some actual experimental data (see Table \ref{tab:results}):
\begin{widetext}
\begin{center}
\begin{table}[ht]
\begin{tabular}{|c|c|c|c|c|}
\hline
\textbf{Material} & $\Delta$ (meV) & $\Delta^*$ (meV) & $T_\mathrm{c}$ Prediction (K) & $T_\mathrm{c}$ from Experiments (K) \\
\hline
\texttt{YBCO} & 25 & 40 & 88 & 93 \\
\texttt{Bi-2212} & 35 & 70 & 98 & 96 \\
\texttt{LSCO} & 15 & 30 & 41 & 38 \\
\texttt{BaFe$_2$(As$_{1-x}$P$_x$)$_2$} & 8 & 12 & 31 & 30 \\
\texttt{FeSe (bulk)} & 2 & 3 & 8 & 8 \\
\hline
\end{tabular}
\caption{Comparison of energy gaps and critical temperatures for selected superconductors.}
\label{tab:results}
\end{table}
\end{center}
\end{widetext}
\begin{enumerate}
\item For \texttt{YBCO} superconductors \cite{YBCO}, typical value for the energy gaps are $\Delta\approx25$ meV and $\Delta^\ast\approx 40$ meV. The theoretical $T_\mathrm{c}$ calculation using Equation (\ref{eq:main}) gives $T_\mathrm{c}\approx 88$ K.
\item For \texttt{Bi-$2212$} superconductors \cite{B}, typical value for the energy gaps are $\Delta\approx35$ meV and $\Delta^\ast\approx70$ meV. The theoretical $T_\mathrm{c}$ calculation gives $T_\mathrm{c}\approx 98$ K.
\item For \texttt{LSCO} single-layer cuprate superconductors \cite{LSCO}, typical value for the energy gaps are $\Delta\approx 15$ meV and $\Delta^\ast\approx 30$ meV. The theoretical $T_\mathrm{c}$ calculation gives $T_\mathrm{c}\approx 41$ K.
\item For \texttt{BaFe${}_2$(As${}_{1-x}$P${}_x$)${}_2$} superconductors \cite{iron}, typical value for the energy gaps are $\Delta\approx 8$ meV and $\Delta^\ast\approx 12$ meV. The theoretical $T_\mathrm{c}$ calculation gives $T_\mathrm{c}\approx 31$ K.
\item For bulk \texttt{FeSe} superconductors \cite{FeSe}, typical value for the energy gaps are $\Delta\approx 2$ meV and $\Delta^\ast\approx 3$ meV. The theoretical $T_\mathrm{c}$ calculation gives $T_\mathrm{c}\approx 8$ K.
\end{enumerate}

The calculation above predicts the gap-to-$T_\mathrm{c}$ ratio in some of the most common superconductors with close agreement.

\section{Doping Dependence of the Critical Temperature}\label{sec:app}

One of the most noticeable features of doped high-temperature superconductors is the emergence of a dome-shaped dependence of the critical temperature $T_\mathrm{c}$ on carrier doping. As charge carriers (typically holes in cuprate systems) are introduced into the parent antiferromagnetic insulator, superconductivity begins to appear at a critical doping threshold. As doping further increases, $T_\mathrm{c}$ rises, and eventually reaches a peak at what is known as the \textit{optimal doping}. Afterwards, it gradually declines in the overdoped regime.

In hole-doped cuprate superconductors, the doping level is typically described by the \textit{hole concentration}, denoted $x$. Both the superconducting energy gap $\Delta$ and the pseudogap $\Delta^\ast$ vary as $x$ changes. To emphasize that, we will now write them explicitly as $\Delta(x)$ and $\Delta^\ast(x)$.

Once the dependence of these gaps is determined, we can apply Equation \ref{eq:main} to express $T_\mathrm{c}$ as a function of $x$:
\begin{align}
T_\mathrm{c}(x)=k_\mathrm{B}^{-1}\frac{\Delta(x)}{1.4+4\log(\Delta^\ast(x)/\Delta(x))}.\label{eq:doped}
\end{align}

When doping level rises, the pseudogap phase is commonly understood to terminate at $x=x^\ast$ where $\Delta^\ast(x^\ast)=\Delta(x^\ast)$ occurs. This condition is often observed at the optimal doping level $x_0$; that is, when $T_\mathrm{c}(x)$ reaches its peak. In certain materials (for instance \texttt{Bi-}$2212$), however, the point where the gaps coincide does not align precisely with the peak of $T_\mathrm{c}(x)$.

In short,
$$x_0\leq x^\ast$$
and the inequality can be strict for certain choices of materials.

\begin{rmk*}
When $x>x^\ast$, the system falls outside of the pseudogap phase regime and it should be understood as $\Delta^\ast(x)=\Delta(x)$ by the notation in this article.

Particularly, in this regime the gap-to-$T_\mathrm{c}$ relation returns to the proportionality formula (\ref{eq:bcs}).
\end{rmk*}

As an application to Equation (\ref{eq:doped}), we can explore how the framework discussed in this article accounts for such behavior and provides a theoretical basis for the observed superconducting dome.

For a concrete example, let us consider a local model where the doping level from $10\%$ to $20\%$ is parametrized by
$$x=0.2-0.1\,t$$
for $0\leq t\leq1$.
Suppose the superconducting and pseudogap energies (in meV) evolve as:
\begin{align*}
\Delta(t) & =10+15\,t\\
\Delta^\ast(t) & =10+15\,t+35\,t^2
\end{align*}
which yields a rough qualitative mirroring of the patterns to experimental data observed in certain cuprate superconductors (cf. \cite{B}).

In this model, the graph for $T_\mathrm{c}(x)$, as shown below, resembles the top of a dome, with the optimal doping level corresponding to $x_0\approx 1.9$, whereas $\Delta^\ast=\Delta$ happens at $x=2$ after the optimal doping $x_0$.

\begin{center}
\begin{tikzpicture}
  \begin{axis}[
    width=7cm,
    height=4cm,
    xlabel={$x$},
    ylabel={$T_\mathrm{c}$ (K)},
    thick,
    grid=major,
    grid style={color=gray!20},
    legend style={
      font=\small,
      at={(-0.2,-0.7)},
      anchor=south west,
      draw=none,
      fill=none,
      cells={anchor=west}
    },
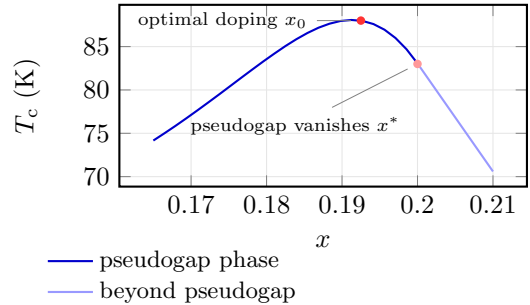
    title={Predicted $T_c(x)$ at different doping levels, $x_0<x^\ast$}
  ]

  \addplot [
    blue!80!black,
    domain=0:0.35,
  ] 
  (
    {0.2 - 0.1 * x},
    { 
      (10 + 15 * x) / ((1.4 + 4 * ln((10 + 35 * x^2+ 15 * x)/(10 + 15 * x))) * 0.086)
    }
  );

  \addplot [
    blue!40,
    domain=-0.1:0,
  ] 
  (
    {0.2 - 0.1 * x},
    { 
      (10 + 15 * x) / ((1.4) * 0.086)
    }
  );

  \addplot[
    only marks,
    mark=*,
    mark size=1.2pt,
    color=red!80
  ] coordinates {(0.1925, 88)};
    \node[pin=180:{\scriptsize optimal doping $x_0$}] at (axis cs:0.1925, 88) {};

  \addplot[
    only marks,
    mark=*,
    mark size=1.2pt,
    color=red!40
  ] coordinates {(0.2, 83)};
    \node[pin=260:{\scriptsize pseudogap vanishes $x^\ast$}] at (axis cs:0.2, 83) {};  

  \legend{pseudogap phase,beyond pseudogap}
  \end{axis}
\end{tikzpicture}
\end{center}

This serves as a numerical demonstration of how the pseudogap phase can persist into the overdoped regime, consistent with experimental observations in particular materials.

On the contrary, if both $\Delta$ and $\Delta^\ast$ are linear in $x$ near the optimal doping level, then $x^\ast=x_0$ would hold.

\section{Discussion on the Fermi-Dirac Distribution}\label{sec:FD}
Even though the derivation in \S\ref{sec:intro} uses the distribution for exclusive fermions, the calculation for free electron gases is analogously valid using the Fermi-Dirac distribution.

The Fermi-Dirac distribution governs free electron gases, and is best suited for electrons in metallic (or other clean materials with negligible between-electron interactions). Those materials are often equipped with a strong conductivity at room temperature. They are also experimentally attested to have worse superconductivity. Nevertheless, we will examine the outcome when such a distribution is applied in our model, and further elaborate on its probable implications.

\subsection{In Two Dimensions}
We replace the distribution function with the Fermi-Dirac distribution and yield
\begin{align}
N&=2A\left(E_\mathrm{c}-\frac{1}{\beta_\mathrm{c}}\log(e^{-\beta_\mathrm{c}E_\mathrm{c}}+1)\right)\label{eq:2DFD}
\end{align}
at $T=T_\mathrm{c}$. The factor $2$ here comes from the spin degeneracy.

We also employ the distribution at $T\rightarrow 0$, which indicates that
$$N=2AE_\mathrm{F},$$
where $E_\mathrm{F}$ is the Fermi energy.

Thus the energy gap is
$$\Delta=-\frac{2}{\beta_\mathrm{c}}\log(e^{-\beta_\mathrm{c}E_\mathrm{c}}+1)\approx-\frac{2}{\beta_\mathrm{c}}e^{-\beta_\mathrm{c}E_\mathrm{c}}<0,$$
which is absurd. Needless to say, it is not a prediction of a negative energy gap; rather, it is a no-go due to the lack of condensation or binding within a free-electron gas. This provides an explanation of why a planar metallic-layer superconductor is simply unlikely to exist.

\subsection{In Three Dimensions}
Things become quite different when we turn to 3-dimensional materials.

The total number of free electron gases (under the density of states $\mathcal{D}(\epsilon)=B\epsilon^{1/2}$ and spin degeneracy 2) is now
\begin{align}
N & =\int_0^\infty2\mathcal{D}(\epsilon)f(\epsilon)d\epsilon\notag\\
& =\int_0^\infty\frac{2B\epsilon^{1/2}}{e^{\beta_\mathrm{c}(\epsilon-E_\mathrm{c})}+1}\notag\\
& \approx\frac{4}{3}BE^{3/2}\left(1-\frac{\pi^2}{8}(\frac{k_\mathrm{B}T_\mathrm{c}}{E_\mathrm{c}})^2\right)\label{eq:NumberFD3D}.
\end{align}
For (\ref{eq:NumberFD3D}), we used the low-temperature expansion and $E_\mathrm{c}\gg k_\mathrm{B}T_\mathrm{c}$.

So we have
$$E_\mathrm{F}^{3/2}\approx E_\mathrm{c}^{3/2}\left(1-\frac{\pi^2}{8}(\frac{k_\mathrm{B}T_\mathrm{c}}{E_\mathrm{c}})^2\right).$$
It follows that
$$E_\mathrm{F}\approx E_\mathrm{c}\left(1+\frac{\pi^2}{12}(\frac{k_\mathrm{B}T_\mathrm{c}}{E_\mathrm{c}})^2\right)$$
and so the gap-to-$T_\mathrm{c}$ relation becomes
\begin{align}
\Delta=2(E_\mathrm{F}-E_\mathrm{c})\approx\frac{\pi^2k_\mathrm{B}^2}{6E_\mathrm{F}}T_\mathrm{c}^2.\label{eq:before_final}
\end{align}
Let us take a closer look at (\ref{eq:before_final}): compared to the proportional relation between $\Delta$ and $T_\mathrm{c}$, which is
$$\Delta^2\propto T_\mathrm{c}^2,$$
it instead says
$$E_\mathrm{F}\cdot\Delta\propto T_\mathrm{c}^2.$$
From this result, we conclude the following:
\begin{enumerate}
\item The fact that $E_\mathrm{F}\gg\Delta$ implies that even if such a material is superconducting, the value of $\Delta$ would typically be small. It hints that Cooper pairs tend to be fragile and unstable in this case, especially since the model used here is solely statistical and ignores the microscopic stability of the Cooper pairs. In fact, compared to other superconductors, most metals (if it has superconductivity to start with) possess a narrow superconducting gap $\Delta$.
\item Nonetheless, these considerations suggest a speculative possibility: if one could devise a mechanism to stabilize Cooper pairs within metallic systems where electrons approximately obey Fermi-Dirac statistics, then such materials might offer a pathway to higher critical temperatures (due to the magnitude of $E_\mathrm{F}$ compared to $\Delta$) from a statistical mechanical point of view.
\end{enumerate}

As a demonstration for the latter point, we plug in some typical values for metallic solids. Say, if the 3-dimensional material with such properties has $\Delta=1.6\text{ meV}$ and $E_\mathrm{F}=7\text{ eV}$, then
$$T_\mathrm{c}\approx300\text{ K},$$
which is desirable.

\section*{Acknowledgments}
The author wishes to express sincere gratitude to Professor Chin-Rong Lee for his invaluable suggestions on the manuscript and for several enlightening and informative conversations that enhanced this work significantly.

\nocite{*}

\bibliographystyle{alpha}
\bibliography{refs}{}

\end{document}